# Better than "Flashes": Transactions

Ruth E. Kastner

Version of 8 February 2011

ABSTRACT. I discuss Gisin's result showing that sets of quantum-correlated spacelike events cannot be described by a covariant probability distribution over hidden variables, and his conclusion that Tumulka's "rGRWf" (relativistic GRW "flash ontology" model) is an appropriate model to resolve the apparent conflict between quantum theory and relativity. The contrasts and similarities between the Transactional Interpretation's (TI) handling of a specific measurement result and that of GRW approaches are clarified. I argue that TI resolves the apparent conflict between quantum theory and relativity described by Gisin in a harmonious manner, without the necessity of modifying the theory.

1. Introduction and Background

This note will argue that the Transactional Interpretation (TI) originated by Cramer (1986) can ably address a perplexing issue raised by Gisin (2010) and compares favorably to the GRW (Girardi, Rimini and Weber, 1986) approaches discussed by him in this context.

GRW approaches were originally undertaken to gain an observer-free account of wave function collapse and thus a solution to the measurement problem. However, it has not been widely recognized that Cramer's Transactional Interpretation (Cramer 1986) already provides an observer-free collapse interpretation without the necessity for any *ad hoc* change to the theory. Moreover, TI provides a straightforward way to resolve the apparent conflict of quantum mechanics with relativity which Gisin discusses in his (2010).

The claim has sometimes been made that GRW has an advantage over TI in that the former spells out a particular measurement result while TI's offer/confirmation wave encounter does not (strictly speaking, the latter determines a *basis* for the determinate outcome while not specifying which one occurs[1]). But this advantage of GRW is only illusory. The GRW outcome is specified by resorting to an *ad hoc* and physically undefined (in terms of any existing theory) 'flash' process. The worst that one can say at present concerning TI is that there is no explicit physical story behind the realization of a particular transaction (outcome) as opposed to a competing "incipient" one, which makes it at least no worse off than GRW in terms of providing concrete physical reasons for a specific measurement result. Meanwhile, TI does give a clear account of the measurement process in terms of absorption. Measurements are performed by setting up situations in which the particle's offer wave (OW) is absorbed at the set of detectors we are interested in, and by discounting runs in which the particle is absorbed (detected) somewhere else. In this regard, a common point of confusion concerning TI is the failure to recognize that confirmation waves (CW) are generated for *all* components of the absorbed offer wave, resulting in a weighted set of incipient transactions corresponding to von Neumann's "Process 1." This set of incipient transactions constitutes an 'ignorance'-type mixture, in that measurement has definitely occurred and the uncertainty concerning outcome is epistemic. The realization of a particular transaction out of a set of incipient ones can be seen as a kind of spontaneous symmetry breaking.[2] So it would not be fair to claim that TI is incapable of providing a clear physical account of measurement.

---
[1] It thus gives a physical basis for von Neumann's "Process 1" or "Projection Postulate."

As a pure interpretation of quantum theory; TI posits no new mathematical structure, and so should not be expected to generate new predictions or to be testable beyond the extent to which basic quantum theory is testable. Rather, it postulates a physical referent for what has been, in traditional usage, uninterpreted[3] mathematical structure in the basic theory: the advanced states appearing in such mathematical components of the theory as Hilbert space inner products, the complex conjugate Schrödinger Equation, and the Born Rule. Specifically, TI proposes that the Born Rule yielding the probability P of outcome X for a system prepared in state $|Y\rangle$, given by

$$P(X|Y) = |\langle X|Y\rangle|^2 \qquad (1)$$

corresponds to the actualization of a transaction between an Offer Wave (OW) described by $|Y\rangle$ and an advanced Confirmation Wave (CW) described by the solution $\langle X|$ to the complex conjugate Schrödinger Equation. The confirmation wave (a necessary but not sufficient condition for a transaction to be actualized, corresponding to an outcome being detected) is emitted upon absorption of a matching OW component by a detector set up to detect systems with property X. This is, for example, a detector in a Stern-Gerlach apparatus which is placed in the path of electrons emerging in the 'spin-up' state for a particular chosen direction. The interpretation of the advanced state $\langle X|$ as a confirmation wave is the physical referent provided by TI for what heretofore has been uninterpreted mathematical machinery in the standard theory. Any outcome in TI thus depends on both the emitter and the absorber(s), and occurs upon the actualization of one transaction (corresponding to 'collapse' along a spacetime interval or intervals) from a set of possible (incipient) ones, each weighted by (1). Note that the actualization of a particular transaction is distinct from the generation of a CW; CWs are emitted for *all* components of an OW which are absorbed.

---

[2] This is being explored in a separate article. The basic idea is that there are many other physical processes, for example the so-called "Higgs mechanism" of the Standard Model of elementary particle theory, in which a set of possible solutions exists but the physical system only adopts one of them, with no determinable reason for it. That situation is certainly not regarded as a 'measurement problem." The heart of the measurement problem of nonrelativistic quantum theory is that no account is given in the usual theory of what counts as a measurement—where the macro/micro 'cut' is, or when the "Projection Postulate" comes into play, and why. TI provides a clear solution to that aspect of the problem from within the theory, while GRW approaches apply a purely *ad hoc* remedy.

[3] By "uninterpreted" here, I mean in mean in an ontological sense, not in the pragmatic sense that, for example, the absolute square of the wave function is to be interpreted as the probability of the associated outcome.

2. Gisin's result

Gisin (2010) has recently argued that under certain conditions, and assuming the traditional understanding of causality (i.e., an event can only influence other events in its future light cone), which will be termed herein "strong causality," Bell's theorem will rule out the ability of all hidden variables (whether local or nonlocal) describable by a covariant probability distribution to reproduce the nonlocal correlations between spacelike detectors for EPR-type entangled states.

Specifically, Gisin considers the usual "Alice and Bob" EPR situation, and defines Alice's and Bob's results $\alpha, \beta$ respectively, as functions $F_{AB}$ [$F_{BA}$] of their measurement settings $\vec{a}, \vec{b}$ and the value of some nonlocal hidden variable $\lambda$. The order of the subscripts on $F$ indicates which measurement is first in the frame considered. Thus if Alice measures first, her outcome $\alpha = F_{AB}(\vec{a}, \lambda)$; if Bob measures first, his outcome $\beta = F_{BA}(\vec{b}, \lambda)$. Gisin then constructs the analogous function $S$ for the outcome measured second, and notes (assuming time-asymmetric strong causality) that it must also be a function of the measurement setting for the first measurement: i.e., $\beta = S_{AB}(\vec{b}, \vec{a}, \lambda)$. Analogous expressions are constructed in the frame in which Bob measures first. Gisin then notes that, if covariance holds, the same $\lambda$ should characterize the results irrespective of the frame considered, so that we must have

$$\alpha = F_{AB}(\vec{a}, \lambda) = S_{BA}(\vec{b}, \vec{a}, \lambda) \qquad (1)$$

and

$$\beta = F_{BA}(\vec{b}, \lambda) = S_{AB}(\vec{a}, \vec{b}, \lambda)) \qquad (2)$$

but there is no $\lambda$ that can satisfy (1) and (2), since they actually imply that $\lambda$ is a local variable and these are already ruled out by Bell's Theorem. Thus, Gisin has ruled out the ability of nonlocal hidden variables to yield a covariant account of actualized outcomes for quantum-correlated spacelike events. This formalizes observations such as Maudlin's (1996) that Bohmian-type "preferred observable" accounts seem to be at odds with relativity.

However, as noted, Gisin's analysis presupposes 'strong causality." That is, it specifies which observer's outcome was prior to the other observer's outcome, with the assumption that the second observer's result depends on the setting and outcome of the first observer. Thus, his result does not rule out the ability of time-symmetric approaches, including those employing time-symmetric hidden variables such as those advocated by Price (e.g., 1997)[4], to yield a covariant account. Nevertheless, I will not be advocating a hidden variable theory here, but rather arguing that the Transactional Interpretation can provide all the benefits of Tumulka's GRW "flash ontology" model, "rGRWf" (2006) without being a modification of the theory.[5]

3. A dilemma re-examined

Tumulka has argued that, in his words, "Either [1] the conventional understanding of relativity is not right, or [2] quantum mechanics is not exact."[6] But this particular dilemma needs to be examined more closely, as horn [1] has more content than is customarily assumed. By [1], Tumulka has in mind the usual assumption that any exact, realist interpretation of quantum theory must involve a preferred inertial frame. But as noted above, there is something more to be questioned in the "conventional understanding" of relativity: an inappropriately strong time-asymmetric causality constraint. So horn [1] really has two different options: [1a] 'there is a preferred frame' or [1b] 'causal influences can be time-symmetric.' Thus option [1] can be chosen *without* embracing a preferred frame, in the form of [1b]. That is, one can reject the necessity of a preferred frame and argue that what is "not right" about the conventional understanding of relativity is the notion that it mistakenly rules out time-symmetric influences.

Whereas GRW "spontaneous localization" approaches such as Tumulka's "rGRWf", in an effort to avoid the preferred foliation that is assumed to be the only option contained in [1], choose [2] and modify quantum theory in an explicitly *ad hoc* manner, TI chooses [1], but not in the sense of [1a] involving a preferred foliation as is usually assumed. Instead, it is noted that relativistic restrictions should be properly considered to apply only to in-principle observable events, and that sub-empirical causal time symmetry--in the sense of our not being constrained to a choice of which of two events is the 'cause' and which the 'effect' --should be accepted via option [1b] (as it is ultimately accepted by Tumulka in relation to his rGRWf in any case; this is discussed below).

Indeed, a similar relaxation of strong causation is just what Tumulka adopts in order to argue that the nonlocal correlations arising between spacelike separated flash events in his model do not violate covariance. He remarks: "An interesting feature of this model's way of reconciling nonlocality with relativity is that the superluminal influences do not have a direction; in other words, it is not defined which of two events influenced the other."[7] Note that, since these are spacelike separated events, there is a frame in which one is first and a different frame in which the other is first, so one could argue that there can be time-reversed causal effects in one frame or the other, depending on which event is arbitrarily considered the "cause" and which the "effect." (One might object here that Tumulka addresses this by saying that no such causal order exists, but that is precisely the case in TI as well; cf. Cramer's account of the EPR experiment (1986, 667-8) .) So we see the relativistic version of GRW already heading in the direction of time symmetry, or at least toward weakening the overly strong "causality" assumption so often presumed in the literature.

---

[4] And see also Evans, P., Price, H., and Wharton, K. (2010) for arguments regarding the advantages of time-symmetric interpretations.
[5] Some researchers apparently view modifications of quantum theory as preferable to interpretations of the unmodified theory. It is this author's view that interpretations of the pure theory are to be preferred if they can provide explanatory power equal to that of modified versions of the theory. While this issue is beyond the scope of the current paper, it is urged that TI not be rejected solely on the basis that it 'does not yield novel predictions' or 'cannot be tested'; such demands are not appropriate for an interpretation of an existing theory that makes no changes to the theory.
[6] *ibid.*

Under TI, sets of possible transactions (whose weights, interpreted as probabilistic propensities, are reflected in the Born Rule), provide a covariant, time-symmetric distribution of possible spacetime events. Moreover, there is nothing about the sets of actualized events in TI that can be seen as noncovariant, as in the actualized events discussed by Gisin. This is because, under TI, it is not assumed that the events (Alice and Bob's outcomes) had a strict temporal causal order. Gisin's observation regarding the non-covariance of actualized events does not apply to sets of actualized events in TI, since all events are dependent on both the emitter's "offer wave" and the absorber(s') "confirmation wave(s)," and, just as in Tumulka's account of his nonlocally correlated flashes, there is no need (nor would it be appropriate) to define which of a set of spacelike separated events is the 'cause' and which is the 'effect' of a particular outcome. The emitter and absorber(s) participate equally and symmetrically in the transaction leading to the outcome(s). Thus, actualized transactions play the part of the "flashes" in Tumulka's model, but without the necessity of modifying the dynamics of quantum theory. While Tumulka has opted for a modification of quantum theory in order to avoid a preferred foliation—our [1a] above—he has also made use of [1b] which, in view of the time-symmetric alternative of TI, obviates the need for modifying quantum theory.

4. Conclusion

Under TI, outcomes arising from actualized transactions constitute a covariant set of spacetime events since they are not constrained by a particular temporal causal order. Thus TI can provide an interpretation of standard quantum theory which is consistent with relativity, without the necessity for ad hoc changes to the theory as in GRW models.

Acknowledgements

The author is most grateful to John G. Cramer, Huw Price, and Avshalom Elitzur for valuable comments.

---

[7] See [5]. preprint version, p. 11

References

Cramer, J. G. (1986). "The Transactional Interpretation of Quantum Mechanics," *Rev. Mod. Phys. 58*, 647-688.

Evans, P., Price, H., and Wharton, K. (2010). "New Slant on the EPR-Bell Experiment," [arXiv:1001.5057v2](arXiv:1001.5057v2) [quant-ph]

Ghirardi, G.C., Rimini, A., Weber, T. (1986). "Unified dynamics for microscopic and macroscopic systems", *Phys. Rev. D 34*, 470–491.

Gisin, N. (2009). "The Free Will Theorem, Stochastic Quantum Dynamics, and True Becoming in Relativistic Quantum Theory,"
http://arxiv.org/PS_cache/arxiv/pdf/1002/1002.1392v1.pdf

Maudlin, T. (2002). *Quantum Nonlocality and Relativity*. Blackwell.

Price, H. (1997). *Time's Arrow and Archimedes' Point*. Oxford.

Tumulka, R. (2006). "Collapse and Relativity," in A. Bassi, D. Duerr, T. Weber and N. Zanghi (eds), *Quantum Mechanics: Are there Quantum Jumps? and On the Present Status of Quantum Mechanics*, AIP Conference Proceedings 844, American Institute of Physics, 340-352 . Presprint version:  http://arxiv.org/PS_cache/quant-ph/pdf/0602/0602208v2.pdf